\documentstyle[12pt]{article}

\def\eg{{\it e.g., }}
\def\ie{{\it i.e., }}

\date{}

\begin{document}

\title{{\LARGE\sf Ground State Structure in a Highly Disordered Spin Glass
Model}}
\author{
{\bf C. M. Newman}\thanks{Partially supported by the
National Science Foundation under grant DMS-92-09053.}\\
{\small \tt newman\,@\,cims.nyu.edu}\\
{\small \sl Courant Institute of Mathematical Sciences}\\
{\small \sl New York University}\\
{\small \sl New York, NY 10012, USA}
\and
{\bf D. L. Stein}\thanks{Partially supported by the
U.S.~Department of Energy under grant DE-FG03-93ER25155.}\\
{\small \tt dls\,@\,physics.arizona.edu}\\
{\small \sl Dept.\ of Physics}\\
{\small \sl University of Arizona}\\
{\small \sl Tucson, AZ 85721, USA}
}
\maketitle

{\bf Abstract:  }  We propose a new Ising spin glass model on ${\bf Z}^d$ of
Edwards-Anderson type, but with highly disordered
coupling magnitudes, in which a greedy algorithm for producing ground
states is exact.  We find that the procedure for determining
(infinite volume) ground states for this model can be related to invasion
percolation with the number of ground states identified as $2^{\cal N}$, where
${\cal N} = {\cal N}(d)$ is the number of
distinct global components in the ``invasion forest''.  We prove that ${\cal
N}(d) = \infty$ if the invasion connectivity
function is square summable.  We argue that the critical dimension
separating  ${\cal N} = 1$ and ${\cal N} = \infty$ is $d_c = 8$. When ${\cal
N}(d) = \infty$, we consider free
or periodic boundary conditions on cubes of side length $L$ and show that
frustration leads
to chaotic $L$ dependence with {\it all} pairs of ground states occuring as
subsequence limits.  We briefly discuss applications of our results to
random walk problems on rugged landscapes.

\section{Introduction}
\label{sec:Intro}

Theoretical work on spin glasses has focused overwhelmingly on the short-ranged
Edwards-Anderson (EA) model \cite{EA} and its infinite-ranged counterpart, the
Sherrington-Kirkpatrick (SK) model \cite{SK}.  The EA Hamiltonian is:
\begin{equation}
\label{eq:Hamiltonian}
{\cal H} = -\sum_{<xy>}J_{xy}\sigma_x\sigma_y\quad ,
\end{equation}
where the sum runs only over nearest-neighbor
pairs of sites on a regular $d$-dimensional lattice.
We will confine
ourselves in this paper to Ising models, \ie the spins $\sigma_x$ take on the
values $\pm 1$.  The couplings $J_{xy}$ are i.i.d. random variables chosen from
a
distribution
symmetric about zero.  One standard choice is a mean zero Gaussian,
though one needn't be confined to that case.

There are a number of open questions \cite{BY} associated with the EA model.
These
include the question of whether it displays a thermodynamic phase transition,
and if
so, in which dimensions; the nature of its large-scale dynamical behavior for
standard local spin-flip dynamics; the relationship between the behavior of
this
model and others, such as the SK model or long-ranged models; and others.  The
question we are concerned with here is the number of ground state pairs the
model
displays in $d$ dimensions.  (Because of the spin-flip symmetry in the
Hamiltonian,
ground states always come in pairs.)
To avoid local degeneracies, we will assume that as in the Gaussian case, the
common
distribution of the $J_{xy}$'s is continuous.

Parisi's analysis \cite{Parisi1}  of the SK model displays the
striking feature of an infinite number of low-temperature phases \cite{Parisi2}
organized in an ultrametric fashion \cite{Mezard}.  A number of workers in the
field have assumed
that a similar result applies also to short-ranged spin glasses, and to other
models with frustration and quenched disorder.

However, an alternative point of view arose in the mid-1980's following a
scaling
{\it ansatz\/} due to MacMillan \cite{Mac}, Bray and Moore \cite{BM}, and
Fisher and
Huse \cite{FH1}.  Their work led to the opposite conclusion \cite{HF}, namely,
that the EA model has only a {\it single\/} pair of ground
states (or of low-temperature pure states) in all finite
dimensions.  Thus, over the past decade two opposing viewpoints have dominated
the
literature.  The first, following the work of the Parisi school, asserts that
short-ranged spin glasses have an infinite number of ground states (in infinite
volume) for all ``nontrivial'' dimensions, \ie $d\ge 2$.  The second,
proceeding
from scaling arguments, argues that such spin glass models possess only a
single pair
of ground states in all finite dimensions.  (Both of the above statements are
normally interpreted to hold for almost every coupling realization.)

In a recent paper \cite{NS2}, hereafter referred to as I, we invented a
model (see also \cite{MCB}) simpler than (but related to) the EA spin glass in
which we could explore these
issues more fully.  The model is analytically tractable, yet complex enough so
that it
displays a surprisingly rich ground state behavior.  We will argue
that our model undergoes a {\it transition in ground state multiplicity\/} at
eight dimensions: below it has a single
pair of ground states, while above it has infinitely (in fact, uncountably)
many.

We are able to demonstrate explicitly the mechanism by which multiple ground
state pairs
arise.  This is the first such demonstration of which we are aware for a
short-ranged spin
glass model in finite dimension.
The nature of the mechanism is related to a
mapping of the ground state structure of our model to invasion
percolation \cite{LB,Chandler,WW}.  We will show that solving the problem of
ground state multiplicity in our
model requires the solution of an interesting problem in invasion percolation.

To summarize, we propose and analyze a tractable short-ranged
finite dimensional spin glass model (although, like
the SK model, it is not physically realistic).  Our techniques enable us to
study the
relationship within the model among quenched disorder, frustration,
dimensionality, and
ground state multiplicity.  We will find in particular that the role of
frustration is
subtle but important, and our model is instructive in illustrating the larger
role
frustration may play in more realistic spin glass models.

The basic features of our model and its analysis have appeared in I.  In this
paper, we
supply a number of arguments omitted for brevity in I, extend our work in
several
directions, and provide proofs for various conclusions drawn in I.  Among these
is a proof
of a theorem (see Section 5) describing conditions under which certain random
growth processes avoid intersection
in $d$ dimensions; as such, its utility extends beyond its particular use in
this
paper.

The outline of the paper is as follows:  In Section 2, we define
our model.  In Section 3 we provide a simple algorithm for finding the ground
state
configuration for any volume with a specified boundary condition, and discuss
some of its
properties.  In Section 4 we demonstrate a second algorithm for finding the
ground state,
thereby mapping our problem onto invasion percolation and showing how the
ground state
multiplicity problem in our model is equivalent to the question of
nonintersection of
invasion regions in invasion percolation.  In Section 5 we provide a heuristic
argument for
the dimension-dependence of the above questions, make part of that argument
rigorous by
relating it to the square-summability of a connectivity function in
invasion percolation, and discuss the dimensionality of the invasion
region as a function of space dimensionality.  This yields the previously
mentioned
transition at eight dimensions.  In Section 6 we discuss the role of
frustration, and point out an important difference between the nature of
multiple ground states in spin glasses and that in random ferromagnets.
Finally, in Section 7 we summarize our results and briefly discuss
their application to some dynamical problems of random walks on rugged
landscapes \cite{NS3,NS4}.

\section{Model}
\label{sec:Model}

We work on a cubic lattice in $d$ dimensions, \ie of sites ${x}\in {\bf Z}^d$
and
edges connecting nearest-neighbor sites only.  The Hamiltonian is of the
standard
Edwards-Anderson Ising form given in Eq.~(\ref{eq:Hamiltonian}). The difference
with the
usual EA Ising model is in the coupling distribution, which now depends on the
system size.
That is, we apply a nonlinear scaling to the couplings which ``spreads them
out''
so much that each
coupling magnitude exists on its own scale --- more precisely, for large enough
system
size each coupling will have at least twice the magnitude of the next smaller
coupling.

We therefore consider a region $\Lambda_L$, which is an $L^d$ cube centered at
the origin.
We achieve the required condition on the couplings by separating their signs
and magnitudes in
the following manner.  Let $\epsilon_{xy}$ be a set of independent identically
distributed (i.i.d.)
symmetric $\pm 1$ valued random variables,
and let $K_{xy}$ be a set of i.i.d.
continuous random variables (\eg uniform on $[0,1]$).
The $\epsilon_{xy}$'s and $K_{xy}$'s are defined on a common probability
space $(\Omega, {\cal F}, P)$ and are independent
of each other.  A point $\omega$ in $\Omega$ may be thought of as a particular
realization of all
$\epsilon_{xy}$'s and $K_{xy}$'s.
Then we set
\begin{equation}
\label{eq:couplings}
J_{xy} = J_{xy}^{(L)} = c_L\epsilon_{xy}\ e^{-\lambda^{(L)}{K_{xy}}}\quad ,
\end{equation}
where $c_L$ is a linear scaling factor which plays no role in ground state
selection and
where the nonlinear scaling factor $\lambda^{(L)}$ is chosen to diverge
fast enough
as $L\to\infty$
to ensure that (with probability one) for all large $L$, each
$|J_{xy}^{(L)}|$ in $\Lambda_L$ is larger than at least twice the next smaller
one.

To see that such a choice of $\lambda^{(L)}$ is possible, note that for any
distinct pair of edges, the function,
\begin{equation}
\begin{array}{ll}
g(\lambda) &= P ( \displaystyle\frac12 \leq e^{-\lambda{ K_{xy}}} / e^{-\lambda
{K_{x^\prime
y^\prime}}} \leq 2)
\\
&= P ( \quad |K_{xy}- K_{x^\prime y^\prime} | \leq \frac{ln 2 }{\lambda}),
\end{array}
\end{equation}
tends to zero as $\lambda \to \infty$ because $K_{xy}$ and $K_{x^\prime
y^\prime}$
are independent {\it continuous} random variables.  The probability
that our desired condition on the $|J_{xy}^{(L)} |$'s in $\Lambda_L$ is
{\it not} satisfied is bounded by
\begin{equation}
\sum_{<xy>}
\sum_{<x^\prime y^\prime >} g ( \lambda^{(L)} )= O ( L^{2d} g (
\lambda^{(L)})),
\end{equation}
where the sums are over edges in $\Lambda_L$.  If we choose
$\lambda^{(L)}$ so that
$g ( \lambda^{(L)}) = O ( L^{-(2d+1+\epsilon)})$ for some $\epsilon >0$,
then the sum of (4) over $L$ is finite, so by the Borel-Cantelli lemma
it follows that
(with probability one) our desired condition will be valid for $L \geq $ some
finite $L^{\ast} (\omega)$.  For example, if the $K_{xy}$'s are
uniform on $[0,1]$, then
$g ( \lambda) = O ( \frac{1}{\lambda})$ and so $\lambda^{(L)}
\geq L^{(2d+1+\epsilon)}$ is a sufficiently fast divergence.

It may be helpful to note that the couplings (2) can be expressed in a simpler
form if
we choose each $\lambda^{(L)}$ to be an {\it odd\/} integer. Letting
$\hat{J}_{xy}$ denote the $L$-{\it independent\/} coupling
$\epsilon_{xy}$exp$(-K_{xy})$, Eq. (2) then becomes
\begin{equation}
J_{xy}^{(L)} = c_{L}(\hat{J}_{xy})^{\lambda^{(L)}}.
\end{equation}
Any continuous symmetric distribution for the $\hat{J}_{xy}$'s (such as
Gaussian) is possible.

We will show in the next section that we have constructed a model whose ground
state
for large $L$
can be found with a type of greedy algorithm.  While this may seem to
suggest that
interesting ground state behavior cannot occur, we shall
see that, surprisingly, this is not
the case.  We point out here, however, that the model has no interesting
behavior at nonzero
temperature.
Its use is only as a means of studying
ground state structure.

\section{Greedy Ground State Algorithm}
\label{sec:GSA}

We begin with a formal definition of infinite-volume ground states for a
specific coupling
realization $\omega$.   We first consider the finite volume $\Lambda_L$  and
for the moment
assume some fixed, boundary condition,
$\hat\sigma$, on  $\partial\Lambda_L$, the boundary of $\Lambda_L$.
We define
$\sigma^*_L$ to be a ground state on $\Lambda_L$
(with boundary condition $\hat\sigma$) if it
minimizes ${\cal H}_L$, where ${\cal H}_L$ is the Hamiltonian of
Eq.~(\ref{eq:Hamiltonian}) with the summation confined to couplings in
$\Lambda_L$
(including those between $\Lambda_L$ and $\partial\Lambda_L$).
The
set of {\it all\/} infinite volume ground states given $\omega$ is then the set
of all
subsequence limits as $L\to\infty$ of $\sigma^*_L$ with some $\hat\sigma_L$
(\ie the set of
all possible limits with all possible boundary conditions).

For the volume $\Lambda_L$ with a fixed boundary condition on
$\partial\Lambda_L$, we now describe an algorithm for finding the ground state
$\sigma^*_L$.
(Non-fixed boundary conditions, such as free and periodic, will be discussed in
Section
\ref{sec:frustration}.)  To do this we rank order the couplings in the
following manner:
the coupling with largest
magnitude
(corresponding to the smallest value of $K_{xy}$)
in $\Lambda_L$ will be said to have rank
one; the next larger coupling will have rank two, and so on.  Now select the
coupling
with rank one; \ie that with the smallest value of $K_{xy}$, and choose the
spins $\sigma_x$
and $\sigma_y$ on its endpoints such that $\sigma_{x}\epsilon_{{x}{y}}\sigma_{
y}>0$, \ie the coupling is satisfied.  Then find the coupling of rank two, and
choose the spins
on its endpoints to satisfy it.  Repeat this procedure, {\it unless\/} a
resulting closed loop
(or path connecting two boundary sites) with previously satisfied couplings
forbids it.  When
that happens, simply proceed to the coupling next in order, and continue until
every coupling
has been tested.

Note that until a cluster of spins
(connected by tested edges) reaches the boundary, only the
relative orientations of the spins in the cluster
are known.  With a fixed boundary condition, the sign
of each spin in a cluster will be determined as soon as it connects to the
boundary.

It is not hard
to see that this algorithm always provides the ground state for any $\Lambda_L$
in which every coupling magnitude is greater than the sum of all those of lower
order in
$\Lambda_L$. This will
be the case for all large $L$ (with probability one) because each coupling
magnitude is
greater than at least twice that of the next smaller one.
To see then that
in the ground state a given coupling {\it must\/} be
satisfied providing this doesn't violate the previously satisfied couplings of
higher rank (or the boundary conditions), consider the clusters formed by
the previously satisfied couplings.  Under the proviso, the two endpoints of
the given
coupling must belong to distinct clusters, at least one of which does not touch
$\partial\Lambda_L$.  If the given coupling were not satisfied in a spin
configuration, then flipping all the spins in that cluster (the one not
touching
$\partial \Lambda_L$) would lower the energy and so the spin configuration
would not be a ground state.

The algorithm outlined here is easily recognizable as simply a version of the
greedy
algorithm --- in effect, we have invented a spin glass model whose exact ground
states
can be found via the greedy algorithm (which for most models is generally a
relatively poor
algorithm for finding ground state configurations or energies).  It is not at
all clear at
this point, however, that the procedure that we outline, when repeated for
ever-increasing
volumes, will have a natural infinite-volume limit --- but we will show in the
next section
that this is in fact the case.  Before we do that, however, we explore some of
the properties
of our model in light of the ground state algorithm just described.

\subsection{Statement of the Problem}
\label{subsec:Statement}

The question of whether this model has multiple
infinite volume ground state pairs is equivalent to whether, as
$L\rightarrow\infty$, a
change in boundary conditions can change a fixed coupling deep in the interior
from being
satisfied to unsatisfied, or vice-versa.

We therefore ask whether any bond $<x_{1}x_{2}>$ exists with
the following property:
for all large $L$,
before any path of satisfied couplings
joining ${x}_1$ and ${x}_2$ within $\Lambda_L$
(i.e., not touching $\partial\Lambda_L$)
is formed according to
the greedy algorithm described above, there already exist two disjoint paths,
one joining ${x}_1$ to the boundary, and
the other joining ${x}_2$ to the boundary.   If such
a bond exists, then whether its coupling is satisfied or unsatisfied will be
determined by the boundary conditions (for all large $L$).

When the boundary of $\Lambda_L$ is sufficiently far from some fixed interior
region $R$, it
may be that no such coupling exists within $R$ ---
each coupling in $R$ is either itself tested before two such disjoint paths can
be
found, or else its endpoints are first connected via some path (not touching
$\partial \Lambda_L$) of
previously tested couplings.  If that is the case
for every finite $R$, then only a single pair of
spin-flip-related ground states exists in the thermodynamic limit.
Otherwise, the system possesses multiple pairs of ground states.

\subsection{``Always Satisfied'' Bonds}
\label{subsec:S1}

One can distinguish between two kinds of satisfied bonds in our model, for a
given coupling
realization $\omega$: there
is the kind which is satisfied in $\omega$ but which would become
{\it unsatisfied\/} in $\omega'$, which is simply $\omega$ with the {\it
sign\/} (\ie the
$\epsilon_{xy}$) of that particular bond reversed.  That is, whether this bond
is
satisfied depends on the specific sign of its corresponding $\epsilon_{xy}$.
The second set
of satisfied bonds are those which are satisfied {\it regardless\/} of their
sign; we
will call them $S1$
bonds (a precise definition is given below).
It is easy to see, for example, that the coupling of highest rank is $S1$.
In fact, each site
in $\Lambda_L$
has at least one $S1$ bond attached to it, namely that whose rank is the
highest of all bonds which connect to that site.  We note that $S1$ bonds
appear in the ordinary
EA model also (\eg any bond whose magnitude is greater than the sum of the
magnitudes
of the adjoining bonds at either of its ends) but they
don't appear to play the crucial role that they play in our model.

Because $S1$ bonds play an important role in what follows, we will devote some
space to study
their properties here.  In particular, it is they which determine the ground
state structure in
our model.  We begin with a precise definition of $S1$
bonds for a given $\Lambda_L$.

\smallskip

$\underline{\rm Definition}$.  A bond will be denoted $S1$ if the following is
true:  its
rank must be greater (\ie its coupling must be of larger magnitude) than at
least one coupling in any path
(not using that bond) connecting its endpoints. In this definition,
we treat all boundary points as automatically connected so that the union of
a path from $x$ to $\partial\Lambda_L$ and a path from $y$ to
$\partial\Lambda_L$
is considered to connect $x$ and $y$. We remark that this notion of connection
within $\Lambda_L$ is a consequence of our dealing with a fixed boundary
condition $\hat\sigma$; when we treat boundary conditions such as free or
periodic (in Section 6), the notion of connection will be modified
accordingly.

\smallskip

According to the definition, an $S1$ coupling is chosen to be satisfied by the
greedy algorithm before any
other path of similarly chosen bonds connects its endpoints.
It is apparent that it is the $S1$ couplings which determine the ground state
configuration.  Satisfied bonds which are not $S1$ (call them $S2$) play no
role in
determining any part of the ground state spin configuration.

We now present some properties of $S1$ bonds which will be useful later.

\noindent 1)  The set of all $S1$ bonds forms a union of trees.  This claim is
obvious from
the definition of $S1$ bonds.

\noindent 2)  The set of all $S1$ bonds spans the set of
all sites in $\Lambda_L$, \ie every site belongs to
at least one $S1$ tree; and furthermore every $S1$ tree touches the boundary of
$\Lambda_L$.

This second claim comes in two parts:  the first has already been shown above
by explicit
construction.  The second part is easily shown by contradiction:  suppose a
given $S1$ tree ``dies''
before reaching the boundary of $\Lambda_L$.  Consider all edges which connect
a point in this
tree to a point not in it.  The coupling of highest rank within this set must
be $S1$.
Therefore, all trees formed of $S1$ couplings reach the boundary.

It may happen, for a given $\omega$, $d$, and $\Lambda_L$, that the $S1$
couplings form either
a single tree or a union of
disjoint trees.
Note that this tree partition of $\Lambda_L$ is the same for all boundary
conditions
$\hat\sigma$. Within each tree the relative sign of the spins is fixed by the
$S1$
couplings; the overall sign for each tree is determined by $\hat\sigma$.
We can now address the question posed earlier ---
different boundary conditions can give rise to
different infinite volume ground states if and only if there
exist fixed neighboring
sites
which belong to {\it disjoint\/} trees of
$\Lambda_L$ for arbitrary large $L$'s.

We are left, however, with several important questions --- how
is the tree partition for
a particular $L$ related to that for some
$L'>L$?  More specifically, how can one be sure that the procedure we
have proposed
has a natural
infinite-volume limit?  In order to
answer these questions, we present an alternative algorithm in the next
section, which will
provide a mapping to invasion percolation.

\section{Invasion Percolation Algorithm}

Before describing our alternative algorithm for obtaining the ground
states, let us note that although the absolute ranks of the $K_{xy}$'s depend
on $L$, the
$K_{xy}$'s
themselves, and hence their relative ordering by rank, does not change with
$L$.
This will allow us to analyze the $L\rightarrow\infty$ limit of our algorithm.

We begin by defining (in all of ${\bf Z}^d$) and for a given $\omega$ (and
hence
a fixed relative ordering of all the
$K_{xy}(\omega)$'s), a growing sequence of trees $T_0 (u), T_1 (u), T_2
(u),\ldots$,
starting from $u\in{\bf Z}^d$, with $T_n (u)$ containing $n$ edges and $n+1$
sites (including $u$).  $T_0 (u)$ consists of $u$ alone, and in general
$T_{n+1} (u)$ is obtained from $T_n(u)$ by considering all edges from sites in
$T_n(u)$
to new sites and adjoining the edge $e_{n+1} (u)=<x_n,x_{n+1}>$ (and new site
$x_{n+1}$) with the smallest value of
$K_{xy}$.  This procedure is identical to that employed in invasion percolation
on ${\bf Z}^d$ (except that we include only edges which connect to new sites).

For a given $L$ and  $u\in\Lambda_L$, let $N_L (u)$ denote the smallest
$n$ such that $T_n(u)$ touches $\partial\Lambda_L$.  The crucial point is that
when every coupling magnitude in $\Lambda_L$ is greater than the sum of all
those of lower order, then for any boundary condition $\hat\sigma$ on
$\partial\Lambda_L$
(and any choice of
$\epsilon_{xy}$'s), in the ground state
$\sigma^{\ast} (\hat\sigma)$, every coupling in $T^{(L)} (u)\equiv T_{N_L
(u)}(u)$
must be satisfied.  To see this,
note that if $J_{{x_n} x_{n+1}}$ were not satisfied (here $n+1 \leq N_L(u)$) in
a given spin configuration, then flipping all the spins in $T_n(u)$ would lower
the energy because $J_{{x_n} x_{n+1}}$ is the coupling on the boundary of
$T_n(u)$ of
largest magnitude.  $\sigma^\ast_u$ is then determined by the tree
$T^{(L)}(u)$, the coupling
signs $\epsilon_{xy}$ on that tree and the boundary condition $\hat\sigma_x$ at
the
boundary site $x$ touched by that tree.

It is clear from the preceding discussion (and that of Section 3) that every
bond
in $T^{(L)}(u)$ (for every $u$) is an $S1$ bond. Since every $T^{(L)}(u)$
touches $\partial\Lambda_L$ and the union of all these edges
(for all $u$'s in $\Lambda_L$) clearly touches every site in $\Lambda_L$, it
must be that this union is the same tree partition
of $\Lambda_L$ as obtained in Section 3 from the union of all $S1$ bonds.
On the other hand, it is clear from the last paragraph, that (for
$u,v\in\Lambda_L$)
the relative sign $\sigma_u^\ast (\hat\sigma ) \sigma_v^\ast (\hat\sigma )$ is
the same
for all choices of $\hat\sigma$ if and only if $T^{(L)} (u)$ and $T^{(L)}(v)$
are
nondisjoint.  Furthermore, one has the following dichotomy concerning the
infinite
volume invasion trees, $T_\infty (u) = \lim_{n\rightarrow\infty} T_n(u) =
\lim_{L\rightarrow\infty} T^{(L)} (u)$:  If $T_\infty (u)$ and
$T_\infty (v)$ are nondisjoint, then $T^{(L)} (u)$ and $T^{(L)} (v)$ are
nondisjoint
for all large $L$; if $T_\infty (u)$ and $T_\infty (v)$ are disjoint, then
$T^{(L)}(u)$
and $T^{(L)}(v)$ are disjoint for all $L$ (such that $u,v\in\Lambda_L$).
We are thus led to the following conclusions, concerning the trees $T_\infty
(u)$ and
their union,
$$
F_\infty = \cup_{u\in {\bf Z}^d} T_\infty (u),
\label{5}
$$
which we call the invasion forest (note that both the trees and the
forest depend only on the $K_{xy}$'s and not the $\epsilon_{xy}$'s):
\begin{description}
\item
(1) In every infinite volume ground state, every coupling in $F_\infty $ is
satisfied.
\item
(2) $F_\infty$ is either a single (infinite) tree or else a union of ${\cal
N}\geq 2$ distinct (infinite)
trees; in either case it spans all of ${\bf Z}^d$.
\item
(3) The former case happens if for every $u,v$, the trees $T_\infty(u)$
and $T_\infty(v)$ intersect.  In this case there is a single infinite
volume ground state pair.
\item
(4) The latter case happens if $T_\infty (u) $ and $T_\infty (v)$ are disjoint
for some $u,v$.  In this case the number of ground state pairs is $2^{{\cal
N}-1}$ and
so is uncountable if ${\cal N}$ is infinite.
\end{description}

In the next section we will discuss the dependence of the value of ${\cal N}$,
and hence of the ground state multiplicity, on the spatial dimension $d$.

\section{Nonintersection in invasion percolation}
\label{sec:Proof}

We have mapped the problem of multiplicity
of ground states in our model to that of whether
invasion percolation has nonintersecting invasion regions.  We can already
answer the question of multiplicity of states of
our model in two dimensions.  Because it is known that
for $d=2$ invasion percolation the trees $T_\infty (u)$
and $T_\infty (v)$ always intersect (in fact are the same modulo
finitely many sites) \cite{CCN}, it follows that our model has only a
single pair of ground states in two dimensions.  Whether any two such
trees in higher dimensions must intersect is an interesting problem
in invasion percolation, which we now consider.

To proceed (mostly nonrigorously), we use a well-known feature of invasion
percolation:  that the invaded
region asymptotically approaches the so-called incipient infinite cluster
(i.e.,  at the
critical percolation probability $p_c$) in the independent bond
percolation problem on the same lattice \cite{Stauffer}.  The fractal
dimension $D$ of the incipient cluster in the independent bond
problem on the $d$-dimensional cubic lattice is known from both
numerical studies and scaling arguments; in particular, $D$ is
dimension-dependent (increasing with d) below six dimensions, but $D=4$ for
$d\ge
6$ \cite{Stauffer}.

The following heuristic argument might then provide an intuitive
picture of our model's behavior.   Consider the infinite-volume invasion trees
$T_\infty (u)$ and $T_\infty (v)$ introduced in the last section.  If each has
a
fractal dimension less than $d/2$, the probability that they will ``miss'' each
other
is greater than zero; if it is greater than $d/2$, they will intersect with
probability one.

This suggests that if the fractal dimension $D_i$ of $T_\infty (u)$ is equal to
$D$,
then the critical dimension of our model is eight.  Below eight dimensions
invasion
regions should always intersect, and hence there would be only one
pair of ground states in our spin glass model; above
that there should be an infinite number.

To make this line of reasoning a bit more precise, let us define a
{\it pair connectedness function\/} $G(y-x)$ as the probability that the site
$y\in T_\infty(x)$.  We can then define $D_i$ by the relation
\cite{Essam,Stauffer}
\begin{equation}
\label{eq:Di}
G(y-x)\sim 1/\|y-x\|^{d-D_i}
\end{equation}
as $\|y-x\|\to\infty$. Note that summing Eq.~(\ref{eq:Di}) over all $y$ in a
box of side length $L$ centered at $x$ yields $L^{D_i}$ as the order of the
(mean) number of
sites in that box which belong to $T_\infty(x)$.

Our main task is then to determine $D_i$ as a function
of space dimension $d$.  First, however, we provide a precise statement of a
condition for nonintersection of invasion trees, which also provides rigorous
justification for part of the heuristic argument presented above.

\bigskip

$\underline{\rm Theorem}$.  If $\sum_{x\in Z^d} G(x)^2<\infty$, then (with
probability one)
there are infinitely many (random) sites $x_1$, $x_2,\ldots$ such that
$T_\infty(x_i)\cap
T_{\infty}(x_j)=\emptyset$.

\medskip

$\underline{\rm Remark}$.  Although we state this theorem in the context of
invasion
percolation, the proof we will now present shows that it remains valid for a
fairly general
class of random growth processes in place of $T_{n}(x)$. The ingredients of the
proof are (statistical) translation and reflection invariance
and the fact that the events $\{T_{n}(x)=A\}$ and $\{T_{n}(y)=B\}$ are
independent as
long as $A$ and $B$ are separated in ${\bf Z}^d$ by some fixed distance. This
fact is
in turn a consequence of the ``local dependence'' of events like
$\{T_{n}(x)=A\}$ on the underlying $K_{xy}$ variables (and the mutual
independence of those
variables).

\medskip

$\underline{\rm Proof}$.  Let $A_n$ denote the event that there exist some $n$
sites,
$x_1,\ldots,x_n$ with $T_\infty(x_i)\cap T_\infty(x_j)=\emptyset$ for $1\le i
<j\le n$.  If
we can show that for every $\epsilon>0$ and every $n$, $P(A_n)\ge1-\epsilon$,
then $P(A_n)=1$ for each $n$ and the desired result follows by letting
$n\to\infty$.  The
desired lower bound on $P(A_n)$ would itself be a consequence of showing that
$P(T_\infty(x)\cap T_\infty(y)=\emptyset)\to 1$ as $\|x-y\|\to\infty$.  To see
this, pick
deterministic $y_1,\ldots,y_n$ with $\|y_i-y_j\|$ large enough, for $i\ne j$,
so that
$P(T_\infty(y_i)\cap T_\infty(y_j)\ne\emptyset)\le\epsilon/{n\choose 2}$; then
\begin{eqnarray}
\label{eq:PAn}
P(A_n)&\ge& P(T_\infty(y_i)\cap T_\infty(y_j)=\emptyset\, \hbox{  for  } 1\le
i<j\le n)\nonumber\\
&\ge& 1-\sum_{1\le i<j\le n}P(T_\infty(y_i)\cap
T_\infty(y_j)\ne\emptyset)\nonumber\\
&\ge& 1-{n\choose 2}\cdot\epsilon/{n\choose 2}=1-\epsilon\quad .
\end{eqnarray}
It remains to obtain a suitable lower bound for $P(T_\infty(x)\cap
T_\infty(y)=\emptyset)$ when
$\|x-y\|$ is large.

Denote by $\rho\left(T_\infty(x),T_\infty(y)\right)$ the minimum Euclidean
distance between
some site in $T_\infty(x)$ and some site in $T_\infty(y)$.  Furthermore, let
$T'_\infty(y)$
denote an invasion region constructed using a completely independent duplicate
set of
variables $\{K'_{xy}\}$ (so that $T_\infty(x)$ and $T'_\infty(y)$ are
independent random
trees).  An elementary but crucial observation is that
\begin{equation}
\label{eq:obs}
P\left(T_\infty(x)\cap T_\infty(y)=\emptyset\right)\ge
P\left(\rho(T_\infty(x),T_\infty(y))>1\right)
=P\left(\rho(T_\infty(x),T'_\infty(y))>1\right)\quad ,
\end{equation}
where the inequality is trivial and the equality follows from a fairly standard
type of
argument (see, e.g.,  Ref.~\cite{AN}) given in the next paragraph which,
roughly speaking, leads to the
conclusion that $T_\infty(x)$ and $T_\infty(y)$ are independent, {\it
conditional\/} on
$\rho\left(T_\infty(x),T_\infty(y)\right)>1$.

To explain more concretely the equality in  Eq.~(\ref{eq:obs}), let us note
that
for any possible configuration $A$ of $T_n(x)$, the event that $T_n(x)=A$
depends only on the $K_{xy}$'s with $<xy>\in\overline{\cal E}(A)$, where
$\overline{\cal E}(A)$ denotes the set of nearest neighbor edges which touch
either one or two vertices of $A$.  If $B$ is a possible configuration for
$T_n(y)$ with $\rho(A,B)>1$ (or equivalently with $\overline{\cal E}(A)\cap
\overline{\cal E}(B)=\emptyset$) then the events $\{T_n(x)=A\}$ and
$\{T_n(y)=B\}$ are
independent since they depend on disjoint sets of the independent
$K_{xy}$'s.  Hence
\begin{eqnarray}
\label{eq:P}
P\left(\rho\Big(T_n(x),T_n(y)\Big)>1\right)&=&\sum_{A,B:\rho(A,B)>1}P\Big(T_n(x)=A,T_n(y)=B\Big)\nonumber\\
&=&\sum_{A,B:\rho(A,B)>1}P\Big(T_n(x)=A\Big)\Big(T_n(y)=B\Big)\nonumber\\
&=&\sum_{A,B:\rho(A,B)>1}P\Big(T_n(x)=A,T_n'(y)=B\Big)\nonumber\\
&=&P\left(\rho\Big(T_n(x),T'_n(y)\Big)>1\right)\ .
\end{eqnarray}
Letting $n\to\infty$ gives the desired equality.

Let $\hat{I}_z$ denote the event that $z\in T_\infty(x)$ and
$\rho(z,T'_\infty(y))\le 1$.
Then $\rho(T_\infty(x),T'_\infty(y))\le 1$ if and only if $\hat{I}_z$ occurs
for some $z$ and hence
\begin{eqnarray}
\label{eq:P2}
P\left(\rho\Big(T_\infty(x),T'_\infty(y)\Big)>1\right)&=&1-P(\cup_{z\in {\bf
Z}^d}\hat{I}_z)\nonumber\\
&\ge&1-\sum_{z\in{\bf Z}^d}P(\hat{I}_z)\nonumber\\
&=&1-\sum_{z\in{\bf Z}^d}P(z\in T_\infty(x))P\left(\rho(z,T'_\infty(y))\le
1\right)\nonumber\\
&\ge&1-\sum_{z\in{\bf Z}^d}G(z-x)\sum_{\|z'-z\|\le 1}G(z'-y)\ .
\end{eqnarray}
By using the reflection invariance of $G(x)$, the last expression can be
rewritten as
\begin{equation}
\label{eq:exp}
1-\sum_{\|w\|\le 1}\left(\sum_{z\in{\bf Z}^d}G(x-y+w-z)G(z)\right)\quad .
\end{equation}

To complete the proof, it clearly suffices to show that
\begin{equation}
\label{eq:suffices}
\Bigl(G\star G\Bigr)(x)\equiv\sum_{z\in{\bf Z}^d}G(x-z)G(z)\to 0\quad \mbox{ as
}\|x\|\to\infty\ .
\end{equation}
But
\begin{equation}
\label{eq:but}
\Bigl(G\star G\Bigr)(x)=\int_{[-\pi,\pi]^d}\left[\hat G(k)\right] ^2e^{-ik\cdot
x}dk
\end{equation}
where
\begin{equation}
\label{eq:where}
\hat G(k)=(2\pi)^{-d/2}\sum_{x\in{\bf Z}^d}G(x)e^{ik\cdot x}\quad .
\end{equation}
Furthermore, $\hat G(k)$ is real since $G(x)=G(-x)$ and so
\begin{equation}
\label{eq:furth}
\int_{[-\pi,\pi]^d}\left[\hat G(k)\right] ^2dk=\int_{[-\pi,\pi]^d}\left|\hat
G(k)\right| ^2 dk
=\sum_{x\in{\bf Z}^d}\left|G(x)\right|^2<\infty\quad ;
\end{equation}
thus $\left[\hat G(k)\right]^2\in L^1\left([-\pi,\pi]^d,dk\right)$ and so by
Eq.~(\ref{eq:but}) and the Riemann-Lebesgue lemma, $\left(G\star G\right)(x)\to
0$ as
$\|x\|\to\infty$, as desired.  $\diamond$

\bigskip

Using this result and Eq.~(\ref{eq:Di}), it follows that the condition
$D_i<d/2$ is sufficient
for nonintersection of invasion regions in $d$ dimensions, and correspondingly,
for our model
to have an uncountable number of ground states.  We now consider this question.

Monte Carlo simulations of invasion percolation on
square and simple cubic lattices \cite{WW} provide strong
evidence that $D_i=D$ in dimensions two and three ($D=91/48$ and
$D=2.53$ \cite{Stauffer}).
In higher dimensions, less is known.  There does
exist, however, an exact solution for invasion percolation on a
Cayley tree \cite{NW}.

One can deduce the fractal dimension $D_i$ of the invasion region using two
different
measures.  One of these is to compute the radius of gyration (i.e., the root of
the mean
square cluster radius) $R$ after the invasion process has completed $n$ steps.
For the Cayley
tree, it was found using Monte Carlo simultations \cite{NW} that, as
$n\to\infty$, $R\sim
n^{1/4}$, consistent with $D_i=4$.

The second measure uses the exact solution mentioned above and is much closer
to
Eq.~(\ref{eq:Di}). The shape function, $S_m^n$, which is the mean number of
invaded
sites on level $m$ of the Cayley tree for an invasion of $n$ steps, is
computed.   By analyzing
Eq.~(9) in Ref.~\cite{NW} (valid for the simplest Cayley tree, i.e., with
coordination number
3) in the limits $\beta\to 1$ and $\alpha\to 1$ (in that order), we find that
$S_m^\infty$ is
proportional to $m$ (with logarithmic corrections) as $m\to\infty$.  The {\it
total\/}
number of sites invaded up to level $m$ thus scales as $m^2$.  The
usual measure of distance on a Cayley tree places level $m$ at distance
${\sqrt m}$ from the origin, leading again to $D_i=4$.

We therefore conclude (nonrigorously) that $D_i=4$ is an {\it upper bound\/}
for the
fractal dimension of an invasion tree on a lattice in finite
dimension.  Given that we expect $D_i =  D$ in any dimension, and given the
known
values of $D$, we conclude that
the critical dimension in our problem is eight.

\section{The Role of Frustration}
\label{sec:frustration}

We have so far argued that our model has a transition in ground state
multiplicity
at eight dimensions.  However, the nature of the ``spin-glassiness'' of our
model, and in
particular the role played by frustration, has not been clarified.  In fact,
suppose that
one were to construct a model of a random {\it ferromagnet\/}  using the
Hamiltonian (\ref{eq:Hamiltonian}) and couplings (\ref{eq:couplings}) but with
all $\epsilon_{xy}=+1$.  For this
system, the ground state in any finite volume with specified boundary
conditions can be
found using the same greedy algorithm described in Section ~\ref{sec:GSA}.  It
is clear
that once again many (infinite volume) ground states (in $d>8$) can be
generated with appropriate choices
of fixed boundary conditions.  (We recall that this is also the case for
uniform ferromagnets in any
dimension or ordinary random ferromagnets \cite{HH} in $d>5$).

So how is the ground state structure of the above random ferromagnet different
from that of our spin glass?  The answer can be found through analyzing the
behavior of
each in the presence of {\it spin-symmetric\/}, coupling-independent boundary
conditions,
such as free or periodic.

In an earlier paper, \cite{NS1} the authors argued that multiplicity of ground
states in
the EA Ising spin glass should be associated with nonexistence of a single
limiting Gibbs
distribution, in the thermodynamic limit, for {\it any\/} coupling-independent
boundary
conditions.  We find that the same association holds here.  While the
conclusion is fairly
clear for any sequence of {\it fixed\/} boundary conditions, in which each
boundary spin
is assigned a definite value, the mechanism for  spin-symmetric (e.g., free or
periodic)
boundary conditions is more subtle.  Its investigation provides deeper insight
into the
nature of our spin glass model, and in particular the role played by
frustration.

We first discuss fixed boundary conditions. Let us now denote the always
satisfied bonds in $\Lambda_L$, as defined in Sec.~\ref{sec:GSA}, as
$S1^w$ bonds, where the $w$ superscript (for ``wired'') is to remind us of
the convention that all boundary points are treated as automatically connected.
(This is analogous to wired boundary conditions in the Fortuin-Kasteleyn
random cluster representation of Ising and Potts models \cite{ACCN}.)
Let us denote by $F_{L}^{w}$ the union of all the $S1^w$ bonds, or
equivalently the union
over $u \in \Lambda_L$ of the invasion trees $T^{(L)}(u)$ (stopped
when they touch $\partial\Lambda_L$). $F_{L}^{w}$ depends on the $K_{xy}$'s
while the overall ``sign'' of any individual tree in $F_{L}^{w}$ (\ie the
sign of a single spin in that tree) is determined finally by the boundary
spin (at the particular site where the tree touches $\partial\Lambda_L$)
and the coupling sign $\epsilon_{xy}$ of the $S1^w$ bond touching that
boundary site. For $d > 8$, as $L$ increases with a given sequence
$\hat\sigma^{(L)}$ of
fixed, coupling-independent  boundary
conditions, the sign of each tree will randomly flip, and no single limiting
ground
state is obtained in the thermodynamic limit.

Indeed the size dependence is such
that {\it all\/} of the uncountably many ground states will appear as limits
along
coupling-dependent subsequences of volumes. To see that, we first condition on
all the
$K_{xy}$'s and let $T_1 , T_2 ,...$ be a list (in some order) of all the
distinct trees from $F_\infty$.
Let $\eta_{i}^{(L)}$ denote the sign of (some fixed $\sigma_{x_i}$ in) $T_i$,
as determined by the $\epsilon_{xy}$'s and the boundary condition
$\hat\sigma^{(L)}$. From the
previous discussion it should be clear that (i) for each $L$, $\eta^{(L)} =
\{\eta_{i}^{(L)}:x_{i} \in \Lambda_{L} \}$ consists of independent random
variables, equally likely to be $+1$ or $-1$, and (ii) for varying $L$, the
$\eta^{(L)}$'s are independent. It follows (for almost every coupling
realization) that
for every $m$ and for every assignment $\overline{\eta}^{[m]}$ of signs
$\overline{\eta}_{i}^{[m]} = \pm 1$ for $i \in \{1,...,m\}$, the collection
$\{L: \eta_{i}^{(L)} = \overline{\eta}_{i}^{[m]} \mbox{ for } i=1,...,m \}$ is
an infinite subsequence
of $L$'s. By a standard subsequence diagonalization argument, it follows that
for every one of the uncountably many assignments $\overline{\eta}$ of signs
$\overline{\eta}_{i} = \pm 1$ for all $i \geq 1$, there exists a subsequence
$\{ L_{j}(\overline{\eta}) : j \geq 1\} $ so that for all $i$,
$\eta_{i}^{(L_{j}(\overline{\eta}))} \to \overline{\eta}_i $ as $j \to \infty$.

Consider now the case of the finite cube $\Lambda_{L} \subset Z^d$, with $d>8$,
and with,
say, free boundary conditions. (We briefly discuss periodic and antiperiodic
b.c.'s later.)  At first it might seem that
the connection between ground states and invasion percolation is no longer
valid, because {\it for every finite volume there is only
a single cluster of always satisfied couplings.\/}  This is because, unlike
when the boundary
spins are fixed, the procedure of satisfying couplings of successively smaller
magnitude will
be continued until all sites are connected. To explain this more fully, we
define a bond
$<xy>$ with {\it both\/} sites $x$ and $y$ in $\Lambda_L$ to be an $S1^f$ bond
exactly as in the
definition of an $S1$ bond in Sec.~\ref{sec:GSA} except that a path is said to
connect
the points $x$ and $y$ only if the path stays entirely within $\Lambda_L$,
never touching the
boundary. In this case, it is not hard to see that the set $F_{L}^f$ of all
$S1^f$ bonds
forms a {\it single\/} tree spanning all of $\Lambda_L$.

To clarify the relation between free b.c. ground states and the invasion
forest $F_{\infty}$,
we begin with a simple but important observation, that every
$S1^w$ bond with both endpoints in $\Lambda_L$ (\ie leaving out any $S1^w$ bond
touching
$\partial\Lambda_L$) is also an $S1^f$ bond. Let us denote by $E_{L}^{f}$ the
set of all
$S1^f$ bonds which are {\it not\/} also $S1^w$ bonds. These are the bonds which
connect
together the distinct trees of $F_{L}^{w}$ to create the single spanning tree
$F_{L}^{f}$.  Note
that every such bond $<xy>$ must have its two endpoints in distinct trees
of $F_{L}^{w}$.  We now make a crucial claim:  as $L\to\infty$, the edges of
$E_{L}^{f}$
will move out to
infinity.  More precisely stated, the intersection of $E_{L}^{f}$ with any
fixed set of bonds
(say, all bonds in $\Lambda_{L_{0}}$ with $L_{0}$ fixed) will, for all large
$L$,
be the empty set.  To see this, first observe that any such edge $<xy>$ not
moving out to
infinity would have two properties:  (a) There would be  no path in $Z^d$
between
$x$ and $y$ with all $K$ values along the path strictly below $K_{xy}$;
 (b) $T_\infty(x)\cap T_\infty(y)=\emptyset$.  It was proved by
Alexander \cite{Alexander} that such edges do not exist in any dimension.

We will next argue that the {\it relative\/} sign between trees flips
randomly,  and so for our spin glass model there is no single limiting pair of
ground
states for a sequence of volumes (chosen independently of the couplings) with
free boundary conditions.  In fact, we will see that in our model, all of
the uncountably many ground state pairs arise via {\it coupling-dependent\/}
subsequences.
This {\it chaotic size dependence\/}, i.e., absence of a limiting ground
state for spin-symmetric, coupling-independent boundary conditions, is
similar to that which
would be found in the EA model were it to possess many ground state
pairs \cite{NS1}.  We argued in Ref.~\cite{NS1} that this is, in fact,
the signature of multiple ground state pairs in spin glasses, and sets
them apart from other systems which may also possess multiple ground
state pairs, such as random ferromagnets.

The argument for random flipping of relative signs between trees in the free
b.c.~case
is like the one used above for flipping of absolute signs of trees in the fixed
b.c.~case, with a few
modifications. Again, we condition on all $K_{xy}$'s and let $T_{1},T_{2},...$
be the distinct
trees of $F_{\infty}$. Now let $\eta_{ij}^{(L)}$ be the relative sign
$\sigma_{x_i}\sigma_{x_j}$ in the free b.c.~ground state.
This is determined (finally)
by the signs
$\epsilon_{xy}$ of the bonds $<xy>$ in $E_{L}^{f}$. Since these bonds move off
to infinity, we
can pick a (sufficiently rapidly growing) subsequence $L_k \to \infty$ so that
the sets
$E_{L_k}^{f}$ are disjoint from each other for different $k$'s. It follows that
the relative signs
$\eta_{1i}^{(L_k)}$ between the first and $i$'th trees will be independent as
both $i$ and $k$ vary.
 Consequently, the same diagonalization procedure as in
the fixed b.c.~case
can be applied to obtain the requisite further subsequences of $L_k$.

Finally, we note that for periodic (or antiperiodic) b.c.'s, $S1^p$ bonds are
defined by
regarding every boundary point as automatically connected to its periodic image
point, but with no other automatic connections. Thus, like in the free
b.c.~case,
every $S1^w$ bond is also an $S1^p$ bond, while the other $S1^p$ bonds must
move to infinity
as $L \to \infty$. This leads to the same conclusions as in the free
b.c.~case.

Let us now contrast this picture with that of the random ferromagnetic
version of our model.  There, a sequence of coupling-independent spin-symmetric
boundary conditions
{\it would\/} yield a single pair of ground states as $L\to\infty$ ---
namely, all spins up and all spins down.
This will be the case in any dimension.  As with the usual models of
ferromagnets and spin glasses,
either of which might have many ground states, the difference from spin
glasses is revealed most sharply
through the use of coupling-independent, spin-symmetric boundary conditions.

\section{Conclusions}

In this paper we have expanded our original discussion in I, supplied
additional proofs and arguments, and looked at our model in greater depth.
In this section we briefly discuss some consequences of our studies,
particularly what (if any) conclusions can be drawn about more realistic
spin glass models, and how our approach can be used to study other
topics, particularly certain dynamical problems.

We first recap our main conclusions, based on both rigorous and nonrigorous
arguments:

\bigskip

$\bullet$ In invasion percolation on ${\bf Z}^d$, the invasion forest
$F_\infty$ has only one tree in dimensions less than eight,
and infinitely many above eight.

$\bullet$ In the spin-glass model introduced in I and discussed here,
there exist only a single pair of ground states below eight
dimensions, and infinitely many above eight.

$\bullet$ Elucidation of the ground state structure exactly
{\it at\/} eight dimensions cannot be addressed by the
techniques provided in this paper, and requires further
study.

$\bullet$ The random ferromagnetic verison of our model
(all $\epsilon_{xy}=1$ in Eq.~(\ref{eq:couplings})) also
has a transition in ground state multiplicity at eight
dimensions.  However, exactly as in the
Edwards-Anderson model, the special feature of
spin-glass ground state multiplicity manifests itself in chaotic
size dependence for spin-symmetric, coupling-independent boundary
conditions, such as periodic or free.

$\bullet$  The special properties of this model render it unsuitable for
drawing firm conclusions about more realistic spin glass models.  However,
it suggests that a third conjecture (besides
those of the Parisi picture and the scaling picture) should be entertained
--- that of a transition in ground state multiplicity at some finite (and,
obviously, greater than one) dimension.

$\bullet$  Our study can resolve one perennially cloudy issue in the
literature; namely, that of whether the joint presence of quenched disorder and
frustration has any universal implications for ground state structure
or multiplicity.  The results presented here make it clear that no
such {\it a priori\/} implications can be drawn.

\bigskip

We close with a brief discussion of applications of our results to other
problems.  The form of the exponential expression in Eq.~(\ref{eq:couplings})
suggests that our
techniques might apply in some way to the problem of a random
walk in a strongly inhomogeneous environment.  If
$K_{xy}$ is thought of as an energy barrier between sites $x$ and
$y$, and $\lambda$ now represents inverse temperature, then exp$(- \lambda
K_{xy})$ may
be interpreted as an Arrhenius factor proportional to
the {\it rate\/}
of making a transition from $x$ to $y$ at fixed
temperature $\lambda^{-1}$.
Random $K_{xy}$'s, like those used throughout this paper, correspond to a
rugged energy
landscape.
We can then prove that, in the limit as
temperature gets small, the order in which sites are visited for
the first time corresponds exactly to the invasion order in invasion
percolation \cite{NS3}.
We further develop conclusions drawn from the theorems proved in
Ref.~\cite{NS3}, and study the nature of broken ergodicity on a system
whose state space resembles a rugged landscape, i.e, has many
metastable states with random fixed barriers separating them \cite{NS4}.
This leads to a surprising degree of emergent structure from
what appears initially as a rather featureless landscape.
The multiplicity of trees in the invasion forest for high dimensions plays an
important role in that work.
We are presently studying how our analysis here and in Refs.~\cite{NS3,NS4} can
be applied in other contexts.


\end{document}